\documentclass[twocolumn,pre,epfs,aps,showpacs]{revtex4}
\usepackage{epsfig}
\begin{document}

\title{Majority-Vote Model on a Random Lattice}

\author{F. W. S. Lima$^{1}$, U. L. Fulco$^{1}$,and R. N. Costa
Filho$^{2}$}
\affiliation{$^{1}$Departamento de F\'{\i}sica,
Universidade Federal do Piau\'{\i}, 57072-970 Teresina,Piau\'\i,
 Brazil \\
$^{2}$Departamento de F\'{\i}sica,
Universidade Federal do Cear\'a, 60451-970 Fortaleza, Cear\'a, Brazil.}

\begin{abstract}

The stationary critical properties of the isotropic majority vote
model on random lattices with quenched connectivity disorder are
calculated by using Monte Carlo simulations and finite size
analysis. The critical exponents $\gamma$ and $\beta$ are found to be
different from those of the Ising and majority vote on the square
lattice model and the critical noise parameter is found to be
$q_{c}=0.117\pm0.005$.

\end{abstract}
\maketitle

\section{Introduction}
Random lattices play an important role in the description of
idealized statistical geometry in a great variety of
fields\cite{Ziman76,Christ82,Moukarzel92,Puhl93}. Besides its
potential to describe the topology in several condensed matter
problems, the comparison of the universality class of systems
between random and regular lattices is also the subject of intense
research\cite{Fisher66,Dotsenko83,Harris74,Espriu86,Janke95}. The
randomness considered in these studies either by random variation
of the coupling strengths, random deletion of bonds or sites, or
by spatially correlated lattices. In particular, according to the
Harris criterion \cite{Harris74} valid for the random-bond
paradigm, random disorder is marginally important to the
two-dimensional Ising model  since the specific heat critical
exponent is $\alpha=0$. However, this criterion could not be
applied to lattices with a non-periodic coordination number. for
these models, Luck\cite{Luck93} formulated a criterion to such
cases.  For example, Janke et al. \cite{Janke94} using the
single-cluster Monte Carlo update algorithm \cite{Wolf89},
reweighting techniques \cite{Swendsen87} and size scaling analysis
\cite{Binder76}, simulate the Ising model in a two-dimensional
Voronoy-Delaunay random lattice. They calculated the critical
exponents and found that this random system belongs to the same
universality class as the pure-two dimensional ferromagnetic Ising
model.

The Voronoi-Delaunay network is a type of disordered lattice
exhibiting a random coordination number that varies from $3$ to
$\infty$, depending on the number of sites. In addition, the
distance $r$ between nearest neighbors changes in a random way
from site to site. Lima et al. used a two-dimensional
Voronoi-Delaunay lattice to study the ferromagnetic Ising model
\cite{Lima00} and the Potts model \cite{Lima00a}. As the bond
length between the first neighbors varies randomly from neighbor
to neighbor, they considered a coupling factor decaying
exponentially as:
\begin{equation}
J_{ij}=J_{0}exp(-a r_{ij})
\end{equation}
where $r_{ij}$ is the relative distance between sites $i$ and $j$,
$J_{0}$ is a constant, and $a \geq 0$ is a model parameter. In
\cite{Lima00} they calculated the critical point exponents
$\gamma/\nu$, $\beta/\nu$ and $\nu$, and concluded that this
random system belongs to the same universality class as the
pure-two dimensional ferromagnetic Ising model. In \cite{Lima00a}
they studied the three-state Potts model and found that critical
exponents $\gamma$ and $\nu$ are different from the respective
exponents of the three-state Potts model on a regular square
lattice. However, a ratio $\gamma/\nu$ remains essentially the
same. They also found numerical evidences that the specific heat
on this random system behave as a power-law for $a=0$ and as a
logarithmic divergence for $a=0.5$ and $a=1.0$.

\begin{figure}
\includegraphics[width=7.5cm]{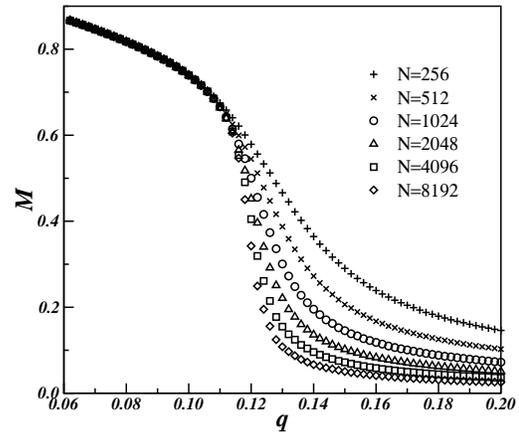}
\caption{Magnetization $m(q)$ as a function of the noise
parameter $q$ for several values of the system size $N$.}
\label{fig1}
\end{figure}

It has been argued that nonequilibrium stochastic spin systems on
regular square  lattice with up-down symmetry fall in the
universality class of the equilibrium Ising model
\cite{Grinstein85}. This conjecture was found in several models
that do not obey detailed balance
\cite{Bennet85,Wang88,Marques90}. Campos et al. \cite{Campos03}
investigated the majority-vote model on small-word network by
rewiring the two-dimensional square lattice. These small-world
networks, aside from presenting quenched disorder, also possess
long-range interactions. They found that the critical exponents
$\gamma/\nu$ and $\beta/\nu$ are different from the Ising model
and depend on the rewiring probability. However, it was not
evident that the exponents change was due to the disordered nature
of the network or due to the presence of long-range interactions.

\begin{figure}
\includegraphics[width=7.5cm]{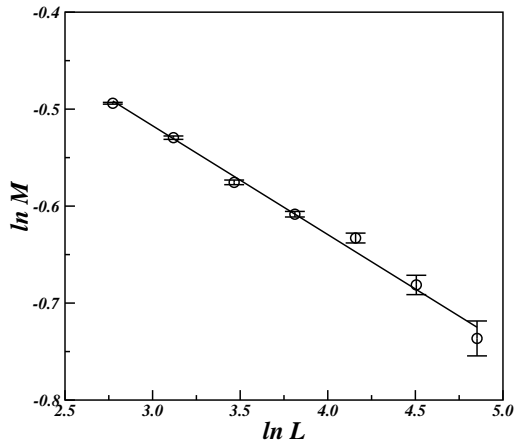}
\caption{Log-log plot of the magnetization at
$q=q_{c}$ versus $L$. The solid line is the best fit with slope
$-\beta/\nu=-0.112(4)$. }
\label{fig2}
\end{figure}

Here, we analyze a relatively simple nonequilibrium model with
up-down symmetry in a random lattice with quenched $conectivity$
disorder ($a=0$), namely the isotropic majority vote model
\cite{Oliveira92}. Our main motivation is to investigate whether
only the presence of quenched lattice disorder is capable of
modifyng the exponents $\gamma/\nu$ and $\beta/\nu$ using the
Voronoi-Delaunay random lattice. Our numerical results suggest
that the critical exponents, in the stationary state, are
different from those of the Ising model and the isotropic majority
vote model on a square lattice. In what follow we will utilize
Monte Carlo simulations and finite-size analysis.

\section{Model and  Simulation}
For each point in a given set of points in a plane, we determine
the polygonal cell that contains the region of space nearest to
that point than any other. We considered two cells neighbors when
they possess an extremity in common. From this Voronoi
tessellation, we can obtain the dual lattice by the following
procedure.

$(a)$ when two cells are neighbors, a link is placed between the
two points located in the cells.

$(b)$ From the links, we obtain the triangulation of space that is
called the Delaunay lattice.

$(c)$ The Delaunay lattice is dual to the Voronoi tessellation in
the sense that its points correspond to cells, links to edges and
triangles to the vertices of the Voronoi tessellation.
\begin{figure}
\includegraphics[width=7.5cm]{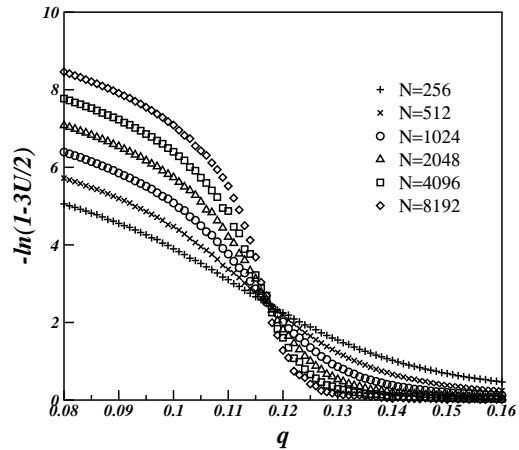}
\caption{Reduced fourth-order cumulant $U(q)$ as
a function of $q$ for several values of $N$. Within the accuracy
of our data, all curves intersect at $q_{c}=0.117(5)$. The value
of $U(q)$ at the intersection is $U^{*}=0.61(2)$.}
\label{fig3}
\end{figure}
We consider a two-dimensional Majority Vote model, on this
Poissonian random lattice, defined \cite{Oliveira92,Mendes98} by a
set of ``voters" or spins variables $\{\sigma_{i}\}$ taking the
values $+1$ or $-1$, situated on every site of a Delaunay lattice
with $N$ sites and periodic boundary conditions, and evolving in
time by single spin-flip like dynamics with a probability $w_{i}$
given by
\begin{equation}
w_{i}(\sigma)=\frac{1}{2}[1-(1-2q)\sigma_{i}S(\sum_{\delta}\sigma_{i+\
delta})]
\end{equation}
where $S(x)=sign(x)$ if $x\neq0$, $S(x)=0$ if $x=0$, and the sum
runs over all nearest neighbors  of $\sigma_{i}$. In this lattice,
the coordination number varies locally between $3$ and $\infty$.
The control parameter $q$ plays the hole of temperature in
equilibrium systems and measures the probability of aligning
antiparallel to the majority of neighbors.
\begin{figure}[h]
\includegraphics[width=7.5cm]{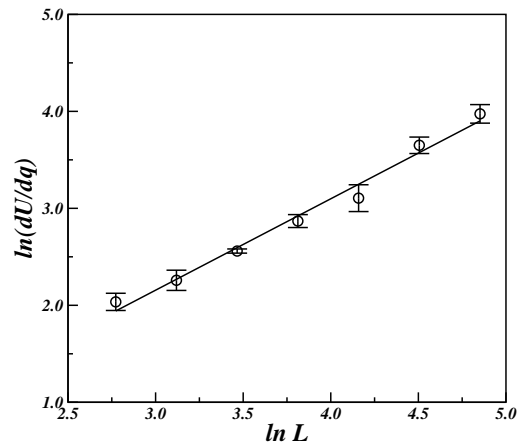}
\caption{Log-log plot of $dU(q)/dq$ at $q=q_{c}$
versus $L$. The solid line is the best fit with slope
$1/\nu=0.94(6)$.}
\label{fig4}
\end{figure}
For simplicity, the length of the system is defined here in terms
of the size of a regular lattice, $L=N^{1/2}$. We perform
simulations for different lattice sizes $N=2^i$, where $i$ varies
from $8$ to $14$. For each size, we generated $50$ randomly chosen
lattice realization, where each simulation started with a random
configuration of spins. From a given configuration, the next one
was obtained as follows: (a) Choose a spin $\{i\}$ at random. (b)
Generate a random number $r$ uniformly distributed between zero
and one. (c) Flip spin $i$ when $r<w_{i}(\sigma)$. In our
simulations, $5$x$10^{4}$ Monte Carlo steps were required for
attaining the stationary state. After that, we estimated the
averages, for any lattice size, using $10^{5}$ Monte Carlo steps.

We define the variable $m=\sum^{N}_{i=1}\sigma_{i}/N$. In
particular, we were interested in the magnetization
\cite{Oliveira92,Binder81}, susceptibility and the reduced
fourth-order cumulant:
\begin{equation}
 m(q)=[<|m|>]_{av}
\end{equation}

\begin{equation}
 \chi(q)=N[<m^{2}>-<|m|>^{2}]_{av}
\end{equation}

\begin{equation}
 U(q)=[1-\frac{<m^{4}>}{3<|m|>^{2}}]_{av}
\end{equation}
where $<...>$ stands for a thermodynamics average and $[...]_{av}$
square brackets for averages over the $50$ realizations. We
calculated the error bars from the fluctuations among
realizations. Note that these errors contain both, the average
thermodynamic error for a given realization and the theoretical
variance for infinitely accurate thermodynamic averages which is
caused by the variation of the quenched random geometry of the
$50$ lattices.

These quantities are functions of the noise parameter $q$ and obey the
finite-size scaling relations
\begin{equation}
 [<|m|>]_{av}=L^{-\beta/\nu}f_{m}(x)[1+...],
\end{equation}
\begin{equation}
 \chi=L^{-\gamma/\nu}f_{\chi}(x)[1+...],
\end{equation}
\begin{equation}
 \frac{dU}{dq}=L^{1/\nu}f_{U}(x)[1+...],
\end{equation}
where $\nu$, $\beta$, and $\gamma$, are the usual critical
exponents, $f_{i}(x)$ are the finite size scaling functions with
\begin{equation}
 x=(q-q_{c})L^{1/\nu}
\end{equation}
being the scaling variable, and the brackets $[1+...]$ indicate
corrections-to-scaling terms.

\section{Results}
\begin{figure}
\includegraphics[width=7.5cm]{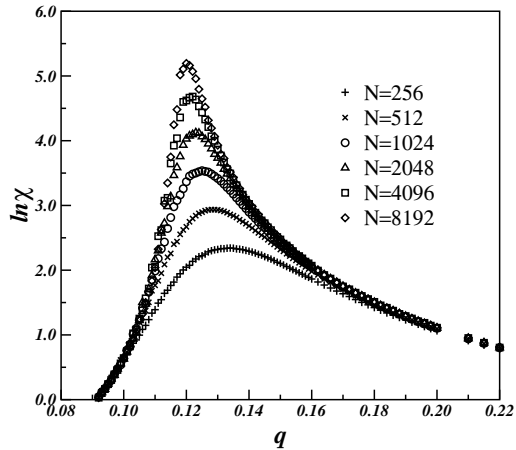}
\caption{Susceptibility $\chi(q)$ as function of
$q$ for several values of $N$.}
\label{fig5}
\end{figure}
In Fig. 1 the magnetization is shown as a function of the noise
parameter $(q)$ for several values of $L$. As can be noticed there
is a phase transition from an ordered state $(M_L>0)$ to a
disordered state $(M_L\approx 0)$. This figure displays that for
$q>q_c$ the magnetization disappears when larger values of $L$ are
considered, whereas it reaches a finite value for $q<q_c$. In
Fig.2 a log-log plot of the magnetization at $q=q_{c}$ versus $L$
is shown. The ratio between the critical exponent
$\beta/\nu=0.112\pm0.004$ is the slope of the straight line fitted
to the data points. Within the numerical accuracy, we found that
these exponents are distinct from the exponents characterizing the
class of universality of the equilibrium Ising model
\cite{Oliveira92}.

To determinate the critical point the reduced fourth-order
cumulant $U$ is plotted against $q$. In Fig. 3, the critical point
$q_{c}$ is estimated when all curves, for different sizes $L$,
intersect in the same point. We get $q_{c}=0.117\pm0.005$ and
$U^{*}=0.61\pm0.02$. The value of $U^{*}$ is, considering the
error bar, the same as the one obtained for Ising model on a
square lattice with periodic boundary conditions
$U^{*}=0.611\pm0.001$. Another way of getting the critical point
is through the relation of scale $q^{\chi}_{max}(L)=
q_{c}+aL^{-1/\nu}$, where we get $q_{c}=0.117\pm0.003$ using
$\nu=1.06$.

To obtain the critical exponent $\nu$, we calculated numerically
$U^{'}(q)=dU(q)/dq$ at the critical point for each value of $L$.
The results are in good agreement with the scaling relation (7).
Then, we plotted $ln U^{'}$ versus $ln L$, as displayed in Fig 4.
The straight line represents the best fit to the data points. The
slope  gives $1/\nu=0.94\pm0.06$, which corresponds to
$\nu=1.06\pm0.08$.

In Fig. 5, we have $ln \chi(q)$ as a function of $q$ for several
values of $L$. In order to study the universality of the model,
the ratio $\gamma/\nu$ was estimated from the log-log plot of the
value of the susceptibility $\chi(q_{c})$ versus $L$. We estimated
the critical exponent $\gamma/\nu=1.51\pm0.04$ from the best fit
of data points, as displayed in Fig. 6.
\begin{figure}
\includegraphics[width=7.5cm]{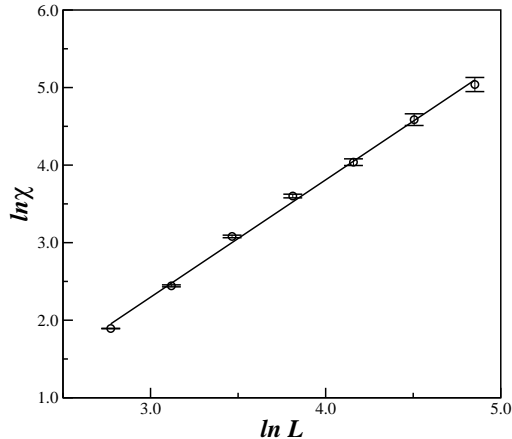}
\caption{Log-log plot of the susceptibility at
$q_{c}$  versus $L$. From it we estimate the critical exponent
$\gamma/\nu=1.51(4)$ as the best fit of data points.}
\label{fig6}
\end{figure}
\section{Conclusion}

We have studied the majority-vote model on Voronoi-Delaunay random
lattices  with periodic boundary conditions. These lattices
possess natural quenched disorder  in their connections. We
verified whether only this type of disorder is significant to
obtain critical exponents different of those found for this model
in the regular lattice (that have the same exponents of the Ising
model in two dimensions). We measure the exponents $\gamma/\nu$,
$\beta/\nu$ and $\nu$. The best fit of these exponents  provided
$\nu=1.06\pm0.08$, $\gamma/\nu=1.51\pm0.04$, and
$\beta/\nu=0.112\pm0.004$ and $U^{*}=0.61\pm0.02$. The critical
exponents $\beta/\nu$ and $\gamma/\nu$ are different from the
exact values of the Ising model and majority-vote model on a
regular square lattice \cite{Oliveira92}. Our results are in
agreement with the results obtained by Campos et al.
\cite{Campos03} that studied this same model on a small world
network presenting quenched disorder and long-range interactions.
They found that the critical exponents depends on the shortcuts
introduced in the network. In summary we showed here that the
presence of quenched connectivity disorder is enough to alter the
exponents $\beta/\nu$ and $\gamma/\nu$ of the pure model and
therefore that disorder is a relevant term to such non-equilibrium
phase-transition. However, the critical value of the forth-order
cumulant remains the same as that of the pure model.

\section{Acknowledgments}

We are grateful to Marcelo Leite Lyra for valuable comments and
suggestions. This work was supported by CNPq, FAPEPI and FAPEAL,
Brazil.

\end{document}